\documentclass{article}

\usepackage{arxiv}

\usepackage[utf8]{inputenc} 
\usepackage[T1]{fontenc}    
\usepackage{hyperref}       
\usepackage{url}            
\usepackage{booktabs}       
\usepackage{amsfonts}       
\usepackage{nicefrac}       
\usepackage{microtype}      
\usepackage{calc}
\usepackage{epsfig}
\usepackage{subcaption}
\usepackage{calc}
\usepackage{amssymb}
\usepackage{amstext}
\usepackage{amsmath}
\usepackage{amsthm}
\usepackage{multicol}
\usepackage{pslatex}
\usepackage{apalike,pifont}
\usepackage{graphicx}
\usepackage{doi}
\usepackage{cite}
\usepackage{amsfonts}
\usepackage{algorithmic}
\usepackage{graphicx}
\usepackage{textcomp}
\usepackage{multirow}
\usepackage{xcolor}
\usepackage{booktabs}
\usepackage{hyperref}
\usepackage{lscape}

\title{A Threat Model for \\ Soft Privacy on Smart Cars}

\author{ \href{https://orcid.org/0000-0002-7045-0213}{\includegraphics[scale=0.06]{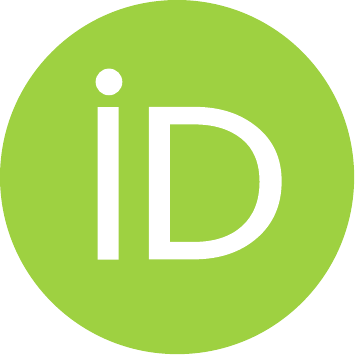}\hspace{1mm}Mario Raciti} \\
	IMT School for Advanced Studies Lucca\\
	Lucca, Italy \\
        Dipartimento di Matematica e Informatica\\
	Università di Catania\\
	Catania, Italy \\
	\texttt{mario.raciti@imtlucca.it} \\
	\And
	\href{https://orcid.org/0000-0002-7615-8643}{\includegraphics[scale=0.06]{orcid.pdf}\hspace{1mm}Giampaolo Bella} \\
	Dipartimento di Matematica e Informatica\\
	Università di Catania\\
	Catania, Italy \\
	\texttt{giamp@dmi.unict.it} \\
}

\renewcommand{\shorttitle}{A Threat Model for Soft Privacy on Smart Cars}

\begin{document}
\maketitle

\begin{abstract}
Modern cars are getting so computerised that ENISA's phrase ``smart cars'' is a perfect fit. The amount of personal data that they process is very large and, yet, increasing. Hence, the need to address citizens' privacy while they drive and, correspondingly, the importance of privacy threat modelling (in support of a respective risk assessment, such as through a Data Protection Impact Assessment). This paper addresses privacy threats by advancing a general modelling methodology and by demonstrating it specifically on \textit{soft privacy}, which ensures citizens' full control on their personal data.  By considering all relevant threat agents, the paper applies the methodology to the specific automotive domain while keeping threats at the same level of detail as ENISA's. The main result beside the modelling methodology consists of both domain-independent and automotive domain-dependent soft privacy threats.
While cybersecurity has been vastly threat-modelled so far, this paper extends the literature 
with a threat model for soft privacy on smart cars, producing 17 domain-independent threats that, associated with 41 domain-specific assets, shape a novel set of domain-dependent threats in automotive.
\end{abstract}


\section{Introduction}
\label{sec:introduction}
Smart Cars~\cite{enisa-report}, Smart Roads~\cite{POMPIGNA2022100986} and Smart Cities~\cite{smart-cities} are revolutionising the automotive domain while promising increased safety, efficiency, and sustainability. 
Such ecosystem generates vast amounts of data that require secure storage, transmission, and processing. Additionally, the integration of various sensors, cameras, and communication systems in modern vehicles creates new opportunities for privacy breaches, raising concerns about data protection measures and corresponding risks. 
Therefore, people’s privacy may be put at stake when they become car drivers. 
The 2023 Global Automotive Consumer Study by Deloitte~\cite{deloitte} finds that consumers may be potentially open to sharing their personal information in exchange for access to vehicle health and maintenance updates, traffic congestion information, and updates for road safety. However, the study also reveals that trust issues, especially in Europe, are hindering consumers from sharing their personal data.

Also, privacy deserves additional scrutiny. While \textit{hard privacy} concerns the various techniques to protect a subject's personal data from everyone else, such as anonymisation and minimisation, \textit{soft privacy} pertains to the range of practices to be followed for the subject to share their personal data with someone else while keeping full control, such as consent mechanisms and impact assessments.
The General Data Protection Regulation (GDPR) is the European answer to the privacy needs of its citizens, and is proving inspirational for similar initiatives worldwide, along the lines of the so called ``Brussels effect''. Our research rests on the observation that privacy issues in the automotive domain are not fully understood at present, although they are certain to demand GDPR compliance. Compliance may be addressed in terms of privacy risk assessment, which in turn demands privacy threat modelling, hence the general motivation for this paper.

\subsection{Context and Motivation}
\label{subsec:context}
Threat modelling is challenging as the analyst faces various problems, such as completeness and threat explosion. On the one hand, completeness may be impactful because failing to account for specific threats would cause pitfalls to the subsequent risk assessment. On the other hand, the research of completeness might also result in threat explosion, which consists of a high number of threats that may not be relevant, feasible, or threats that are redundant. Completeness and redundancy are considered by Raciti and Bella~\cite{vehits23} through their systematic approach that leverages the LINDDUN state-of-the-art privacy threat modelling framework~\cite{landuyt2020} to elicit privacy threats for the automotive domain. The final list of threats online~\cite{repo} is produced by taking a domain-dependent approach and by embracing the threats from various sources, including in particular ENISA's ``Good practices for security of smart cars''~\cite{enisa-report}. 

In this context, the motivations for the present paper are manyfold:
\begin{itemize}
    \item ENISA's ``Good practices for security of smart cars'' is among the most relevant sources about car cybersecurity in Europe. However, its treatment of privacy is very limited, as we shall detail below, hence a deeper close-up is necessary.
    \item The privacy assessment by Raciti and Bella~\cite{vehits23} is very relevant. However, as the authors state, it is affected by subjectivity, hence the need for a more algorithmic approach to privacy threat elicitation, such as one of combinatoric nature.
    \item Hard and soft privacy, as noted above, distil out the two main spheres of privacy and demand dedicated scrutiny. While LINDDUN~\cite{landuyt2020} considers them to some extent, it does not specifically tailor them to the automotive domain.
    \item Cybersecurity and privacy are certain to intertwine but are often confused with one another. As we shall see below, existing efforts to distil them out, specifically in the automotive domain, are very few and worth expanding.
\end{itemize}

\subsection{Research Question and Contributions}
Following the context and motivations given above, this paper focuses on soft privacy in the automotive domain from the threat modelling perspective. It does so, by stating the following research question:
\begin{quote}
RQ \textit{What are the soft privacy threats for the automotive domain and how to elicit them?}
\end{quote}
It may be observed that the analyst may want to adopt a domain-dependent or domain-independent approach. The research question sets the scene for the former, meaning that it aims at threats that are specific for the identified target domain, i.e., automotive in this case.

This paper answers the research question by proposing an innovative methodology for privacy threat modelling. It applies a combinatoric approach that has two benefits. The first is the elicitation of domain-independent threats from analysing one or more sources from the state of the art. The second is the elicitation of domain-dependent threats, derived from the combination of a generic threat knowledge base and domain-specific assets. It also leverages LINDDUN~\cite{Deng2011}, which has become the de-facto standard privacy threat modelling framework and is inherently domain-independent. It adopts the mentioned ENISA report as a source of specific and comprehensive knowledge on the automotive domain.

Beside the methodology just outlined, this paper contributes a list of 17 soft privacy threats that are domain-independent but are also combined with 41 specific assets of the automotive domain, so as to produce a set of domain-dependent soft privacy threats for the automotive domain.
As a first additional by-product, the paper also proposes the new list of domain-independent threats as a possible extension of the generic threat knowledge base provided by LINDDUN. A second positive side effect is the suggestion of domain-dependent threats as possible candidates to extend the existent threat taxonomy proposed by ENISA. In fact, a merge of the soft privacy threats with such a taxonomy provides a novel, comprehensive security-and-privacy-embracing threat taxonomy. In particular, this compensates for the lack of privacy-specific threats in the ENISA report, hence paving the way to a better understanding of the potential privacy risks associated with the automotive domain.

\subsection{Article Summary}
\label{sec:article-summary}
The organisation of the manuscript follows a simple waterfall style. Section~\ref{sec:related-work} outlines the related work and Section~\ref{sec:background} gives an overview of LINDDUN. Section~\ref{sec:methodology} describes our privacy threat modelling methodology. Section~\ref{subsec:demo} demonstrates the methodology by applying it to the automotive domain along with a case study, and Section~\ref{sec:conclusions} concludes.

\section{Related Work}
\label{sec:related-work}
The challenges implicated by threat modelling led Wuyts et al.~\cite{8844639} to highlight the problems of current knowledge bases, such as limited semantics and lack of instantiating logic. Also, the authors discussed the requirements for a privacy threat knowledge base that streamlines threat elicitation efforts.

Furthermore, it is also noteworthy to recall that the process of threat modelling inherently implies assumptions and arbitrary decisions. Landuyt et al.~\cite{landuyt2020} highlighted the influence of assumptions to the outcomes of the analysis during the risk assessment process, more precisely in the threat modelling phase in the context of a LINDDUN privacy threat elicitation.

In addition, several attempts were made for the purposes of threat modelling in the automotive domain. Vasenev et al.~\cite{vehits19} were among the first to apply an extended version of STRIDE~\cite{stride} and LINDDUN~\cite{Deng2011} to conduct a threat analysis on security and privacy threats in the automotive domain. In particular, the case study is specific to long term support scenarios for over-the-air updates. Moreover, this work suggests that the privacy topic in the automotive domain has not reached the same level of maturity as cybersecurity.

In general, threat modelling is part of a wider process, that is risk assessment. Wang et al.~\cite{Wang2021} proposed a threat-oriented risk assessment framework tailored for the automotive domain, with the aim, among the others, of overcoming assumptions and subjectivity. This framework can be considered a precursor to ISO/IEC:21434\cite{iso21434}, which was defined a year later. Also, the authors applied STRIDE and the attack tree method for the threat modelling.

Moreover, Chah et al.~\cite{CHAH202236} applied the LINDDUN methodology to elicit and analyse privacy requirements of CAV system, while respecting the privacy properties set by the GDPR. Such attempt represents a solid baseline for the broader process of privacy risk assessment tailored for the automotive domain.
Finally, Bella et al.~\cite{Bella2023} advanced a dedicated risk assessment framework for privacy risks in modern cars. They proposed a double assessment, combining an asset-oriented ISO approach with a threat-oriented STRIDE approach.

The above works addressed crucial topics such as threat elicitation, threat knowledge base, privacy threat analysis and privacy risk assessment, both in general and specifically tailored to the automotive domain. However, on the best of our knowledge, there are no other works proposing a privacy threat modelling methodology that features the application of a combinatoric approach to elicit both domain-independent and domain-dependent soft privacy threats.

\section{A Primer on LINDDUN}
\label{sec:background}
It is convenient to provide a brief introduction to LINDDUN before proceeding with the description of the methodology.
LINDDUN is a privacy threat modelling methodology, inspired by STRIDE, that supports analysts in the systematical elicitation and mitigation of privacy threats in software architectures. LINDDUN privacy knowledge support represents one of the main strength of this methodology, and it is structured according to the 7 privacy threat categories encapsulated within LINDDUN's acronym~\cite{Deng2011}:

\begin{itemize}
    \item\textit{Linkability} An adversary is able to link two items of interest without knowing the identity of the data subject(s) involved.
    \item\textit{Identifiability} An adversary is able to identify a data subject from a set of data subjects through an item of interest.
    \item\textit{Non-repudiation} The data subject is unable to deny a claim (e.g., having performed an action, or sent a request).
    \item\textit{Detectability} An adversary is able to distinguish whether an item of interest about a data subject exists or not, regardless of being able to read the contents itself.
    \item\textit{Disclosure of information} An adversary is able to learn the content of an item of interest about a data subject.
    \item\textit{Unawareness} The data subject is unaware of the collection, processing, storage, or sharing activities (and corresponding purposes) of the data subject’s personal data.
    \item\textit{Non-compliance} The processing, storage, or handling of personal data is not compliant with legislation, regulation, and/or policy.
\end{itemize}

The framework considers the state-of-the-art privacy properties, according to the terminology introduced by Pfitzmann~\cite{anon_terminology}. These are categorised as hard privacy and soft privacy properties. In particular, unlinkability, anonymity and pseudonymity, plausible deniability, undetectability and unobservability, and confidentiality (hiding data content, including access  control) are under the umbrella of hard privacy; user content awareness (including feedback for user privacy awareness, data update and expire) together with policy and consent compliance are, on the other hand, soft privacy properties.

The LINDDUN framework provides a set of threats specific to privacy, named as ``threat catalogue'', in the form of threat trees. These privacy threat trees are inspired by the Secure Development Lifecycle (SDL)~\cite{sdlc} and reflect common attack patterns~\cite{linddun-nutshell} on the basis of state-of-the-art privacy developments, structured according to LINDDUN or STRIDE threat category and Data Flow Diagram (DFD) element type.
Such trees provide a formal way to describe the security of systems based on a variety of attacks. Basically, the root node represents the ultimate goal, e.g., the threatening to a property, the children nodes embody different ways of achieving that goal, namely refinements, hence leaves represent basic-level attacks that can not be further refined. In addition, non-leaf nodes can be conjunctive (logic AND) or disjunctive (logic OR)~\cite{schneier1999attack}.

\begin{figure}[ht]
    \centering
\includegraphics[width=0.75\textwidth]{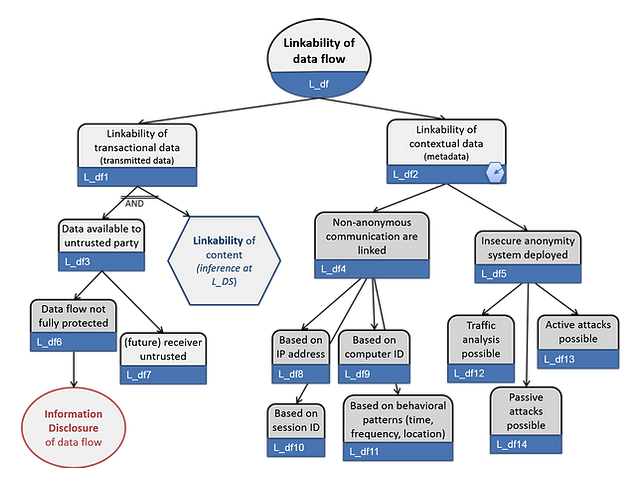}
    \caption{Example of a LINDDUN threat tree: Linkability of Data Flow.}
    \label{fig:linddun-tree-example}
\end{figure}

An example tree is presented in Figure~\ref{fig:linddun-tree-example} for the Linkability of Data Flow, which can be achieved through $L\__{df1}$ ``Linkability of the contextual data (transmitted data)'', e.g., IP address necessary for communication, and $L\__{df2}$ ``Linkability of the transactional data (metadata)'', that is the actual data transmitted. Both of these provide various attack paths which are not necessarily limited to the LINDDUN property analysed, namely Linkability of data flow could lead to the Information Disclosure of data flow if we consider the following path, assuming the Linkability of content, namely inference, at data store applies:
$$L\__{df1} \to L\__{df3} \land L\__{DS} \to L\__{df6} \to ID\__{df}$$

Despite LINDDUN threat trees may lack some semantics and have minimal selection criteria to express potential threats~\cite{8844639}, they still aim at providing a valuable overview of potential threat types that seeks to be general, hence suitable for a privacy threat analysis of any application domain. Moreover, the application of LINDDUN may lead to a high number of threats that may not be relevant, feasible, or important, thereby being labor-intensive and time-consuming, especially for complex or large systems. Hence, the advantage of having a catalogue of privacy threats, which are broad and applicable to various domains, may result in the problem of threat explosion.

\section{A Privacy Threat Modelling Methodology}
\label{sec:methodology}
Privacy is a complex and multifaceted concept that may be interpreted in different ways in different contexts, yet in the first place it is a fundamental human right. In fact, NIST defined it as \textit{the right of a party to maintain control over and confidentiality of information about itself}~\cite{nist-privacy}. In a GDPR fashion, we may summarise privacy as the right of an individual, that is, the data subject, to control or influence what information related to them may be collected, processed and stored, and by whom and to whom that information may be disclosed.

Privacy is frequently intertwined with security, as privacy concerns frequently arise in relation to security matters. While threat modelling has traditionally been approached from a security perspective, it should be emphasised that privacy and security are two distinct concepts, hence they cannot be used interchangeably. In fact, the analyst should not rely on risk assessment models that are only security-centric, as these may not be sufficient to capture all aspects of privacy risks. A challenge for all privacy threat modelling approaches is how to consider the impact on data subjects involved in the privacy threat. This aspect is stressed in law and regulations compliance, i.e., in the Data Protection Impact Assessment (DPIA), required under the GDPR, to help identify, assess, and mitigate privacy risks associated with data processing activities. From a data subject's perspective, a DPIA is an important aspect of privacy threat modelling, as it ensures that organizations consider the potential impact of their data processing on individuals' privacy rights and take appropriate measures to address any risks.
Arguably, a DPIA would benefit from a privacy threat model.

We advance a privacy threat modelling methodology that incorporates both domain-independent and domain-specific knowledge and considers the potential consequences on the privacy of individuals as its cornerstone. The pivotal notion that our methodology relies upon is the use of a combinatoric approach to elicit both domain-independent threats and domain-dependent threats. In particular, the former embody a generic threat knowledge base that consists of what is already known at present, whilst the domain-specific threats are derived from the first. Furthermore, our methodology identifies at least four variables that contribute to model privacy threats, i.e., the specific privacy property, the threat agents, the application domain and the level of detail. 
The combinatoric approach as well as each of the identified variables are discussed below.

\subsection{The Combinatoric Approach}
\label{subsec:combinatoric-approach}
 The approach consists of three steps:

\begin{enumerate}
    \item Domain-Independent Threat Elicitation
    \item Domain-Dependent Asset Collection
    \item Domain-Dependent Threat Elicitation
\end{enumerate}

The first step involves the collection of domain-independent threats from relevant sources.
The second step consists of the collection of a list of assets for the target domain from relevant sources.
The third and last step aims at producing a list of domain-specific threats. In particular, for each domain-independent threat elicited in Step 1, this step associates the assets enumerated in Step 2, hence obtaining a list of domain-dependent threats. The sheer association expresses the object of the threat that was domain-independent in the first place, thereby making it domain-dependent.

For the sake of convenience, the steps are noted through the creation of a table that should contain at least the following columns: Source, Threat, Assets. In particular, the source reflects the origin document or catalogue where the given threat was derived from, whereas the threat and assets columns are self-explanatory.

It is noteworthy that the subjectivity through this approach reduces to the collections to be done through the first two steps. Prose is not meant to be re-interpreted and modified by the analyst following step 3. By contrast, the sibling approach~\cite{vehits23} declares its subjectivity in the specific wording of each threat.

\subsection{The Specific Privacy Property}
\label{subsec:property}
Privacy relates to the control that individuals have over their personal information, including how it is collected, used, and shared. According to the state of the art~\cite{danezis,Deng2011}, we can distinguish between two degrees of privacy, namely hard privacy and soft privacy. We identify in such properties the first element of scrutiny to build a privacy threat model, and contend that each of them deserves a specialised treatment.

\textit{Hard privacy} refers to data minimisation, based on the assumption that personal data is not disclosed to third parties. The threat model includes service provider, data holder, and adversarial environment, where strategic adversaries with certain resources are motivated to breach privacy, similar to security systems~\cite{Deng2011}. Examples of hard privacy include anonymisation of data, data minimisation, and data retention policies, including the use of algorithms such as k-anonymisation, t-closeness, differential privacy. For instance, a company that collects user data may anonymise the data before sharing it with third parties, ensuring that the users' identities remain protected. Similarly, a company may limit the amount of data it collects, processes, or stores to a minimum, and may have policies in place for deletion once processing that data is no longer necessary.

\textit{Soft privacy}, on the contrary, is based on the assumption that the data subject lost control of their personal data and has to trust the honesty and competence of data controllers~\cite{Deng2011}. Examples of soft privacy include transparency and consent mechanisms, data subject access rights, and privacy impact assessments. For example, a company may provide clear and concise privacy notices to inform users about how their data will be used and shared. They may also obtain users' consent before using their data for purposes beyond the original scope. Additionally, companies may, and should to be GDPR-compliant, provide users with the right to access, modify, or delete their personal data.

In summary, while hard privacy focuses on minimising the risks associated with the collection and retention of personal data, soft privacy focuses on the appropriate use and sharing of personal data while respecting individuals' rights to control their data.
It is clear that, in addition to hard privacy and soft privacy, \textit{cybersecurity} plays a major, complementary role in terms of protection against the unauthorised access of data.

Our methodology pays specific attention to both incarnations of privacy, but is described only on soft privacy in this paper due to space constraints.

\subsection{The Threat Agents}
\label{subsec:threat-agent}
Threat agents are individuals, groups, or systems that have the capability to exploit vulnerabilities and cause harm to a system or organisation. In threat modelling, understanding the capabilities, motivations, and objectives of threat agents is crucial for identifying and prioritising threats. There are different types of threat agents, including insiders, outsiders, script kiddies, hacktivists, cybercriminals, and nation-state actors.

In the context of privacy threat modelling, we refer to a threat agent as any entity, individual or group, who poses a threat to an individual's privacy. Unlike the security literature, which traditionally refers to such entities as ``adversaries'' or ``attackers'', here the term threat agent also includes other sources of risks to privacy, as a threat agent is less security-connotated and not limited to malicious actors only. In fact, we also consider three additional actors directly from GDPR, i.e., data controller, data processor, and third party as threat agents. 
Therefore, a threat agent can be one or more of the following entities:

\begin{itemize}
    \item \textit{Attacker} Anyone, including an insider, or anything, including malware, acting with malicious intent to compromise a system to breach users' privacy.
    \item \textit{Data controller} The natural or legal person, public authority, agency or other body which, alone or jointly with others, determines the purposes and means of the processing of personal data;
    \item \textit{Data processor} A natural or legal person, public authority, agency or other body which processes personal data on behalf of the controller;
    \item \textit{Third party} A natural or legal person, public authority, agency or body other than the data subject, controller, processor and persons who, under the direct authority of the controller or processor, are authorised to process personal data.
\end{itemize}

This taxonomy allows us to better model the data subject's perspective during the threat modelling exercise. For instance, it might be relevant to understand whether to classify a threat against data as a data breach or data leak~\cite{enisa-threat-landscape}.
A data breach is defined as \textit{any breach of security leading to the accidental or unlawful destruction,
loss, alteration or unauthorised disclosure of or access to personal data transmitted, stored or otherwise processed}
(article 4.12 GDPR), while a data leak is an event that can cause the unintentional release of sensitive, confidential or protected data due to, for example, misconfigurations, vulnerabilities or human errors.
Many use these notions interchangeably, yet they entail fundamentally different concepts that mostly lie in how they happen and what is the agent causing them.
Data breach may be understood as an intentional attack brought by an attacker with the goal of gaining unauthorised access and the release of sensitive, confidential or protected data. In other words, a data breach is a deliberate and forceful attack against a system or organisation with the intention of stealing data. On the other hand, a data leak may be understood as not including intentional attacks, and it may be attributed to, e.g., data controller and data processor.

Our methodology assumes that one or a combination of multiple threat agents may be involved in the threat scenario.

\subsection{The Application Domain}
\label{subsec:domain-independent-vs-domain-dependent}
The application domain in threat modelling identifies two prevailing approaches, i.e., domain-dependent and domain-independent. Domain-dependent threat modelling is specific to a particular application domain, such as healthcare, finance, or automotive, and it takes into account the unique characteristics of the domain itself, thus it may be more accurate and effective. On the other hand, domain-independent threat modelling is not specific to any application domain and can be applied to a wide range of systems. This approach uses generic threat categories, such as spoofing, tampering, and repudiation in STRIDE, to identify and prioritise threats. A generic threat knowledge base comes particularly useful in situations where there is no prior knowledge of the system or domain. LINDDUN, for example, is a domain-independent methodology.

A combination of the two approaches may offer a more effective and efficient analysis, picking the advantages of both. In addition, we may obtain the same results from a domain-dependent approach by starting from a generic threat knowledge base and associating it to domain-specific characteristics, i.e., assets. This is what, for instance, ENISA did in the study reported in ``Good practices for security of smart cars''~\cite{enisa-report}. They addressed the domain-dependent vs domain-independent dilemma by taking the combinatoric approach we mentioned above. Our methodology inherits the combinatoric approach, i.e., the combination of domain-independent threats with domain-dependent assets to elicit domain-dependent threats.

\subsection{The Level of Detail}
\label{subsec:level-of-detail}
The last element of scrutiny derives from the level of detail --- of the statement describing a threat. For example, ``Unchanged default password'' is certain to be more detailed than (the more abstract) threat ``Human error''. Normally, the analyst strives to choose a consistent level of detail till the end of the exercise.
The level of detail becomes relevant in the context of threat modelling and, subsequently, in risk assessment exercises with respect to the likelihood estimation of a threat.

The concepts of hyponym and hypernym are semantic relations between terms with respect to their hierarchy. Both play an important role in understanding the level of detail in a statement. In particular, hypernyms are broader, more general terms that encompass a category or set of items, whilst hyponyms are more specific, narrower terms that belong to a category or set.
The relationship is asymmetric, meaning that while a hypernym can include many hyponyms, that is, a hypernym is an umbrella term, a hyponym can only have one hypernym.

A \textit{higher level of detail} (hyponym) implies the analyst is able to estimate the likelihood of such a threat with more precision. However, an excessive level of detail leads to the degeneration of the threat to a ``measurable event'', hence to an exact assignment of the likelihood, that is either the bottom or the top in the given range.
If the analyst's aim is to obtain a checklist of measurable events, a higher level of details represents the best option.
However, the most appropriate level of detail, that is, the choice of employing hypernyms or hypernyms, should be considered within the main picture, and the analyst 
will choose it with some inevitable bias.

Our methodology addresses this variable by keeping the level of detail of the source knowledge base unaltered. This choice does not introduce subjectivity.

\section{Demonstration in the Automotive Domain}
\label{subsec:demo}
We apply the privacy threat modelling methodology described above to address the research question. In particular, we set the variables discussed through Section~\ref{sec:methodology} as follows:

\begin{itemize}
    \item[P] Soft Privacy
    \item[T] Attacker, Data Controller/Processor, Third Party
    \item[D] Domain-Dependent
    \item[L] Abstract
\end{itemize}

\subsection{Domain-Independent Threat Elicitation}
\begin{figure}[ht]
    \centering
\includegraphics[width=0.75\textwidth]{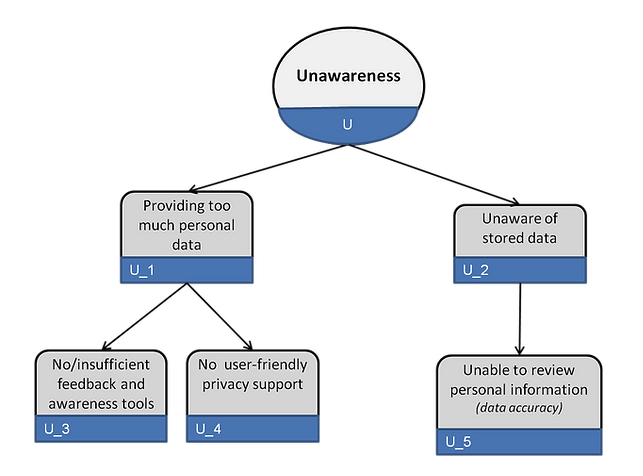}
    \caption{LINDDUN threat tree: Unawareness.}
    \label{fig:unawareness}
\end{figure}

Soft privacy is the target property, therefore we must consider the LINDDUN threats that refer to such property, i.e., U(nawareness) and N(on-compliance). For each node of the U and N property trees, we annotate the pertaining threat in a table. For the sake of demonstration, Figure~\ref{fig:unawareness} illustrates the threat tree for the U(nawareness) property.

As one of the aims in eliciting the list of soft privacy threats is completeness, we may also want to extend the list of domain-independent threats by adding two other sources. In particular, we include the 8 threats that were found by Raciti and Bella~\cite{vehits23} to be outstanding with respect to LINDDUN. In detail, they account for the 2 threats from the ENISA report that fall under the ``Legal'' category, i.e., ``Failure to meet contractual requirements'' and ``Violation of rules and regulations/Breach of legislation/
Abuse of personal data'', and 6 threats from the ``Calculation of the complete Privacy Risks list 
v2.0''~\cite{owasp} document, i.e.,  ``Consent-related issues'', ``Inability of user to access and modify data'', ``Insufficient data breach response'', ``Misleading content'', ``Secondary use'', ``Sharing, transfer or processing through 3rd party''. These threats relate to soft privacy as per the definition of soft privacy we covered previously in Section~\ref{subsec:property}. It is noteworthy to mention that this addition is still possible without consequences on the domain variable, as such threats are general privacy threats, i.e., they ignore domain specific entities. Hence, such threats can be analysed in relation with (virtually) any application domain. Table~\ref{tab:step-1} illustrates our results, which represent a first by-product of this paper as generic, domain-independent soft privacy threats, following this step.

\begin{table}[ht]
\caption{Domain-independent soft privacy threats elicited in Step 1.}\label{tab:step-1} \centering
\begin{tabular}{|c|l|}
\hline
\textbf{Source} & \textbf{Threat}   \\ \hline
                                 & Providing too much personal data                                                                                      \\ \cline{2-2} 
                                 & Unaware of stored data                                                                                                \\ \cline{2-2} 
                                 & No/insufficient feedback and awareness tools                                                                          \\ \cline{2-2} 
                                 & No user-friendly privacy support                                                                                      \\ \cline{2-2} 
\multirow{-5}{*}{U}              & Unable to review personal information (data accuracy)                                                                 \\ \hline
                                 & \begin{tabular}[c]{@{}l@{}}Attacker tampering with privacy policies and makes\\ consents inconsistent\end{tabular}    \\ \cline{2-2} 
                                 & Incorrect or insufficient privacy policies                                                                            \\ \cline{2-2} 
                                 & Inconsistent/insufficient policy management                                                                           \\ \cline{2-2} 
\multirow{-4}{*}{N}              & Insufficient notice                                                                                                   \\ \hline
                                 & Failure to meet contractual requirements                                                                              \\ \cline{2-2} 
\multirow{-2}{*}{ENISA}          & \begin{tabular}[c]{@{}l@{}}Violation of rules and regulations/Breach of legislation/\\ Abuse of personal\end{tabular} \\ \hline
                                 & Consent-related issues                                                                                                \\ \cline{2-2} 
                                 & Inability of user to access and modify data                                                                           \\ \cline{2-2} 
                                 & Insufficient data breach response                                                                                     \\ \cline{2-2} 
                                 & Misleading content                                                                                                    \\ \cline{2-2} 
                                 & Secondary use                                                                                                         \\ \cline{2-2} 
\multirow{-6}{*}{OWASP}          & Sharing, transfer or processing through 3rd party                                                                     \\ \hline
\end{tabular}
\end{table}

It is convenient to provide a brief and generic explanation of the threats, referring to the descriptions provided by their sources. In particular, U(nawareness) means being unaware of the consequences of sharing information. Users are often not aware of the impact of sharing data on social networks or with other services. Lack of feedback and awareness tools can cause users to be unaware of the information they are sharing. It is important that privacy support is user-friendly and that users have the ability to modify privacy settings. Data subjects are often unaware of what data a system has collected and stored about them, so they should always have the ability to review their own data.

Furthermore, N(on-compliance) means not being compliant with legislation, regulations, and corporate policies. Non-compliance can lead to fines, loss of image and credibility. Users should have control over their data and be able to decide who has access to it through consent. Privacy policies and consents need to be stored correctly and consistently to ensure compliance. Lack of user-friendly policy management and insufficient implementation can lead to non-compliance with user consent requirements. Clear and transparent notice of applied policies is necessary to inform users about data collection, storage, and processing.

In addition, the ``Legal'' threat category by ENISA highlights that not following contractual requirements can result in negative consequences such as financial loss, safety issues, privacy breaches, and operational disruptions. Moreover, non-compliance with international or European regulations and laws, e.g., GDPR and UNECE regulations, may have a significant impact on users' privacy, such as the unauthorised exposure of their personal information.

Finally, the threats originating from OWASP raise concerns about legal, technological and organisational aspects that focus on real-life risks. These include: collection of data without the user's consent or aggregating consent; the lack of ability for users to change or delete incorrect data relating to them; not informing the affected persons (data subjects) about a possible breach or data leak; user misled through confusing, ambiguous, or misleading content or descriptions of their privacy practices; operators' use personal data in ways unrelated to the original purpose of the collection, without even informing affected people of the change; providing user data to any third-party, without obtaining the user’s consent, and/or lack of clarity about policies or protocols employed by the third party.

\subsection{Domain-Dependent Asset Collection}
\begin{figure}[ht]
    \centering
\includegraphics[width=0.5\textwidth]{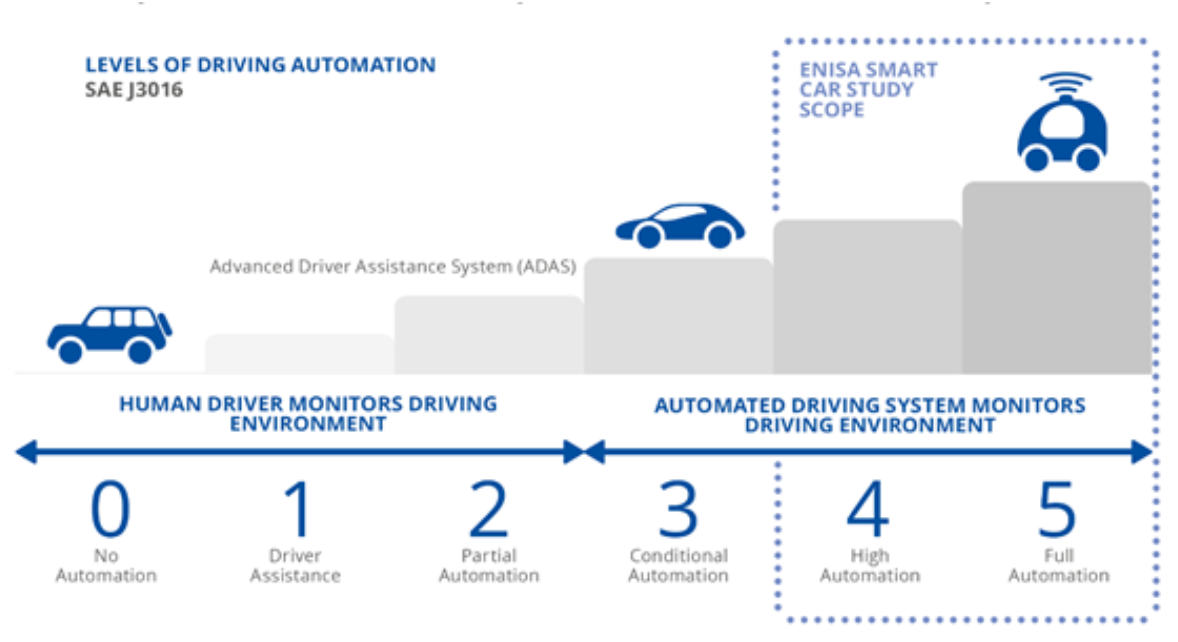}
    \caption{SAE vehicles automation levels as defined in SAE J3016. Source: ENISA}
    \label{fig:sae}
\end{figure}

For Step 2, we leverage two sources from the state of the art, i.e., the assets identified in the work proposed by Bella et al.~\cite{Bella2023} and the ENISA taxonomy  of the key assets in the automotive domain. 
The former presents the following list of assets: ``Personally Identifiable Information'', ``Special categories of personal data'', ``Driver’s behaviour'', ``User preferences'', ``Purchase information'', ``Smartphone data'', ``GPS data'', ``Vehicle information'', ``Vehicle maintenance data'', ``Vehicle sensor data''.
The report by ENISA focuses on Automated Driving System-Dedicated Vehicle (ADS-DS)~\cite{sae}, i.e., semi-autonomous and autonomous cars, and V2X communications. Figure~\ref{fig:sae} depicts such SAE vehicles automation levels. The focus of the study is on smart cars that, as connected systems, have the necessary capabilities to autonomously perform all driving functions under certain (or all) conditions, and are able to communicate with their surroundings including other vehicles, pedestrians and Road-Side Units (RSU). Moreover, the key concepts analysed by ENISA do not only concern passenger cars but also commercial vehicles (e.g. buses, coaches, etc.), including self-driving, ride-sharing vehicles that can be shared with other users.

The assets proposed by ENISA are categorised in: ``Car sensors and actuators'', ``Decision Making Algorithms'', ``Vehicle Functions'', ``Software management'', ``Inside vehicle Communication Components'', ``Communication Networks and Protocols'', ``Nearby External Components'', ``Network and Domain Isolation Features'', Servers'', ``Systems and Cloud Computing'', ``Information'', ``Humans'', ``Mobile Devices''. For the sake of brevity, we only quote the descriptions of the assets under the ``Information'' category, which are inherently most likely to be affected by soft privacy threats:

\begin{itemize}
    \item\textit{Sensors data} This asset refers to data that is gathered by the
different smart car sensors and which will be transmitted to the appropriate ECU for processing.
    \item\textit{Keys and certificates} This asset refers to the different keys and certificates used for security purposes (such as authentication, securing the exchanges, secure boot, etc.). Keys are stored in devices embedded in the vehicle (e.g. ECU) and/or in servers depending on their use.
    \item\textit{Map data} This asset refers to the information about the car environment. Map data allows us to increase the passenger safety by correlating its information with the sensor data. Contrary to GNSS, which gives only information about the geolocalisation, map data gives information about the surrounding environment.
    \item\textit{V2X information} This asset refers to the various information exchanged via V2X communications (e.g. emergency vehicle approaching, roadworks/collision warning and traffic information).
    \item\textit{Device information} This asset refers to the various information related to a device embedded in a smart car (e.g. ECU, TCU) or connected devices (e.g. smartphones, tablet). This includes information such as type, configuration, firmware version, status, etc., of different smart car sensors and which will be transmitted to the appropriate ECU for processing.
    \item\textit{User information} This asset refers to smart cars user (e.g. driver, passenger, etc.) information such as name, role, privileges and permissions.
\end{itemize}

Moreover, soft privacy is part of privacy, which is related to security, thereby all the assets proposed by ENISA may be virtually involved in the execution of the combinatoric approach.

During the execution of this step, we identified some redundancy between the two sources, that is, some assets proposed by Bella et al. are \textit{embraceable} with the ENISA taxonomy. Thereby, we explicitly picked the following assets from the paper contribution: ``Special categories of personal data'', ``Driver's behaviour'', ``User preferences'', ``Purchase information''. The remaining assets, according to our scrutiny, are already contemplated in ENISA taxonomy.

At this point, one may argue that also threats from Bella et al. could be considered, with respect to the previous step. However, those threats were mapped to hard privacy properties in Raciti and Bella's contribution, hence they are not relevant for the target of this paper, that is, soft privacy. Yet, they would be significant for an exercise tailored to that particular property.

\subsection{Domain-Dependent Threat Elicitation}
In the last step, we associate the results of the previous steps. For each domain-independent threat elicited in Step 1, we assign the assets from Step 2 that we deem to be potentially affected by that particular threat. In general, a threat may apply to multiple assets, therefore for some threat-asset pairs we annotate multiple ENISA assets or, in case all assets are affected, we add the label ``All assets'' for the sake of brevity. In particular, most assets that we deem to be potentially affected by the threats fall under the ENISA category ``Information''.

While Table~\ref{tab:results} summarises the results, which represent the second by-product of this paper, following the last step, we may present an exemplification of each domain-dependent threat, with the additional aim of providing the reason behind the various threat-asset(s) associations:

\begin{itemize}
\item \textit{Providing too much personal data} Users may be required to provide an excessive amount of personal data to access certain features on the infotainment system of the vehicle or to use a connected service, potentially putting their privacy at risk.
\item \textit{Unaware of stored data} Users may not be aware of what data is being collected and stored by the onboard systems of their car, leading to potential privacy violations.
\item \textit{No/insufficient feedback and awareness tools} OEM and/or car-related services may not provide adequate feedback or awareness tools to inform users about what data is being collected and how it is being used, leading to potential privacy violations.
\item \textit{No user-friendly privacy support} OEM and/or car-related services may lack user-friendly privacy support, such as clear and accessible privacy policies or easy-to-use privacy controls, leaving users unsure about how to protect their privacy.
\item \textit{Unable to review personal information (data accuracy)} Users may be unable to review or correct their personal information that is being collected and stored by the onboard systems of the car, potentially leading to inaccurate data being used for decision-making.
\item \textit{Attacker tampering with privacy policies and makes consents inconsistent} Attackers may attempt to tamper with privacy policies of the car or manipulate a user's consent, leading to potential privacy violations.
\item \textit{Incorrect or insufficient privacy policies} OEM and/or car-related services may have privacy policies that are incorrect or insufficient, leaving users unaware of what data is being collected and how it is being used.
\item \textit{Inconsistent/insufficient policy management} OEM and/or car-related services may not have effective policy management processes in place, leading to potential privacy violations.
\item \textit{Insufficient notice} OEM and/or car-related services may not provide sufficient notice to users about what data is being collected and how it is being used, potentially leading to privacy violations.
\item \textit{Failure to meet contractual requirements} OEM and/or car-related services may not meet the contractual privacy requirements agreed upon with users, leading to potential privacy violations.
\item \textit{Violation of rules and regulations/Breach of legislation/Abuse of personal data} OEM and/or car-related services may violate rules and regulations or abuse personal data, potentially leading to privacy violations.
\item \textit{Consent-related issues} Users may not fully understand what they are consenting to when using connected services or providing personal data to their car's onboard systems, potentially leading to privacy violations.
\item \textit{Inability of user to access and modify data} Users may be unable to access or modify their personal information that is being collected and stored by their car's onboard systems, potentially leading to privacy violations.
\item \textit{Insufficient data breach response} OEM and/or car-related services may not have effective processes in place for responding to data breaches, potentially leading to privacy violations.
\item \textit{Misleading content} OEM and/or car-related services may provide misleading content to users about what data is being collected and how it is being used, potentially leading to privacy violations.
\item \textit{Secondary use} Personal data collected by OEM and/or car-related services may be used for secondary purposes without the user's consent, potentially leading to privacy violations.
\item \textit{Sharing, transfer or processing through 3rd party} Personal data collected by OEM and/or car-related services may be shared, transferred or processed by third parties without the user's consent or knowledge, potentially leading to privacy violations.
\end{itemize}

\subsection{Case Study}
\label{subsec:case-study}
This section presents a case study that relies on the latest breaking news and articles about privacy incidents in the automotive domain. In particular, we employ classical web searches as a source of relevant information by building queries as ``privacy automotive'', ``automotive breach'', ``smart car privacy'', et similia, in the \textit{News} search filter offered by Google. If we matched some news with a soft privacy threat from the previous exercise, then we would be able to give some statistics about the occurrences of such threat, hence inferring an estimation of its likelihood. For the sake of brevity, we only present some illustrative examples of news that matched with one or more of the proposed soft privacy threats.

A data breach at Toyota Motor's Indian business~\cite{reuters} might have exposed some customers' personal information. ``Toyota Kirloskar Motor (TKM) has been notified by one of its service providers of an incident that might have exposed personal information of some of TKM’s customers on the internet''. This piece of news perfectly embodies two threats that we find in Table~\ref{tab:results}, i.e., ``Insufficient data breach response'' and ``Violation of rules and regulations/Breach of legislation/Abuse of personal data'', because of GDPR compliance.

Furthermore, we also find another news that represents the threat ``Violation of rules and regulations/Breach of legislation/Abuse of personal data''. The Dutch Data Protection Authority (DPA) investigated Tesla's camera-based ``Sentry Mode'' security system~\cite{automotive_news}, which is designed to protect the vehicle against theft or vandalism while it is parked. It does this by taking footage with four cameras on the outside of the vehicle.
This specific threat has now received a mitigation measure from the manufacturer, as the company altered security cameras to be more privacy-friendly and avoid GDPR violations. Originally, when Sentry Mode was enabled, this system was on by default. The cameras continuously filmed everything around a parked Tesla and stored one hour of footage each time.

In addition, we also found a review~\cite{cnn} that perfectly matched with the implications related to the threat ``No user-friendly privacy support''. The article discusses a suggestion for a new feature to be added to the Ring Car Cam. The author proposes an Alexa-based voice command that would temporarily turn off the interior camera and microphone. This suggestion is based on the author's wife's volunteer work, which involves discussing private and privileged information about children's legal cases on the phone. The author's wife currently uses the physical privacy shutter to prevent the camera from recording video and audio inside the car. However, she sometimes forgets to flip the shutter up or down. Therefore, the author proposes a hands-free privacy trigger that would allow the user to enable or disable privacy mode with a voice command. This feature would eliminate the need for the user to physically manipulate the shutter, making it easier to maintain privacy while driving.

\subsection{Findings}
\label{subsec:findings}
In this section, we discuss the findings from the previous experiment. The application of the combinatoric method to the automotive domain yielded notable results, which are available in Table~\ref{tab:results}, as stated above. In particular, we produced a novel, refined list of soft privacy threats that are domain-dependent. In fact, if we associate the generic threat knowledge base pertaining to soft privacy, collected at the end of Step 1, with the automotive-specific assets collected at the end of Step 2, we obtain domain-specific soft privacy threats for modern cars with a homogeneous level of detail and dependent on the automotive domain. This answers the research question.
A confirmation of the concreteness and relevance of such threats for the automotive domain was proven, although informally, with the employment of web searches.
Furthermore, the ENISA report is more tailored to security, in terms of the \textit{property} variable, thereby the new list of threats may be taken as possible candidates to expand the ENISA report, hence enriching the broader threat knowledge base in the automotive domain over soft privacy. 
Moreover, we cannot claim that no more valid candidates exist, yet the list of threats results complete with respect to the state-of-the-art privacy threat knowledge base, i.e., LINDDUN threat catalogue and automotive asset taxonomy, i.e., ENISA. Nonetheless, these threats remain valid candidates for the international community’s evaluation.

\section{Conclusions}
\label{sec:conclusions}
This paper faced the challenge of privacy threat modelling focusing specifically on soft privacy and on the automotive domain, as its research question states.
The research question found an answer in the advancement of a threat modelling methodology that is general for privacy, as it revolves around (at least) four variables. Moreover, while the methodology ultimately aims at eliciting domain-dependent threats, the combinatoric approach it takes has two additional benefits. One is that, by collecting threats from relevant albeit scattered sources, it does not change their level of detail hence it does not introduce (additional) subjectivity. The other one is that, by taking a combinatoric approach through threat elicitation, it produces a set of domain-independent threats but, at the same time, a set of automotive domain-dependent threats.

The methodology is then demonstrated on soft privacy, yielding 17 domain-independent threats, which can be tailored to 41 assets to become automotive domain-dependent. We argue that soft privacy has not received such an in-depth treatment thus far, considering that our set of threats is appreciably larger than both LINDDUN's and ENISA's, and it clearly introduces more facets of soft privacy. 
A natural follow-up of our work is to analyse the application of the combinatoric approach proposed by our methodology to different tuples of variables, for example addressing hard privacy at a very high level of detail. 

In particular, to achieve a fuller account on the level of detail during the elicitation process, we expect to be able to leverage modern, intelligent techniques from the area of Natural Language Processing. For example, we are currently looking at the concepts of hypernym and hyponym towards the definition of a metrics to assist the analyst through the choice of a consistent level of detail for threat descriptions.
An interpretation of the findings of this paper is that (soft) privacy is finally threat-modelled as extensively as it deserves and, in particular, as cybersecurity traditionally has been. In consequence, the risks for ``\textit{natural persons with regard to the processing of personal data and on the free
movement of such data}''~\cite{gdpr}, particularly when those natural persons drive smart cars, can be now assessed more precisely than before. 

\begin{landscape}

\begin{table*}[ht]
\centering
\caption{Results of the application of our methodology to the automotive domain: list of soft privacy threats.}\label{tab:results}
\begin{tabular}{|c|l|l|}
\hline
\textbf{Source}     & \textbf{Threat}                                                                           & \textbf{Assets}                       \\ \hline
\multirow{5}{*}{U}           & Providing too much personal data                                                                                           & User information, Special categories of personal data                                                                                                                                                                                            \\ \cline{2-3} 
                             & Unaware of stored data                                                                                                     & \begin{tabular}[c]{@{}l@{}}Map data, V2X information, Device information, User information,\\ Special categories of personal data, User preferences, Purchase information\end{tabular}                                                           \\ \cline{2-3} 
                             & No/insufficient feedback and awareness tools                                                                               & \begin{tabular}[c]{@{}l@{}}Map data, Device information, User information,\\ Special categories of personal data, Driver's behaviour, User preferences,\\ Purchase information\end{tabular}                                                      \\ \cline{2-3} 
                             & No user-friendly privacy support                                                                                           & \begin{tabular}[c]{@{}l@{}}Sensors data, Map data, V2X information, Device information,\\ User information, Special categories of personal data, Driver's behaviour,\\ User preferences, Purchase information\end{tabular}                       \\ \cline{2-3} 
                             & Unable to review personal information (data accuracy)                                                                      & User information, Special categories of personal data                                                                                                                                                                                            \\ \hline
\multirow{4}{*}{N}           & \begin{tabular}[c]{@{}l@{}}Attacker tampering with privacy policies and\\ makes consents inconsistent\end{tabular}         & \begin{tabular}[c]{@{}l@{}}Sensors data, Key and certificates, Map data, V2X information,\\ Device information, User information, Special categories of personal data,\\ Driver's behaviour, User preferences, Purchase information\end{tabular} \\ \cline{2-3} 
                             & Incorrect or insufficient privacy policies                                                                                 & All assets                                                                                                                                                                                                                                       \\ \cline{2-3} 
                             & Inconsistent/insufficient policy management                                                                                & All assets                                                                                                                                                                                                                                       \\ \cline{2-3} 
                             & Insufficient notice                                                                                                        & \begin{tabular}[c]{@{}l@{}}Sensors data, Key and certificates, Map data, V2X information,\\ Device information, User information, Special categories of personal data\end{tabular}                                                               \\ \hline
\multirow{2}{*}{ENISA}       & Failure to meet contractual requirements                                                                                   & All assets                                                                                                                                                                                                                                       \\ \cline{2-3} 
                             & \begin{tabular}[c]{@{}l@{}}Violation of rules and regulations/Breach of legislation/\\ Abuse of personal data\end{tabular} & All assets                                                                                                                                                                                                                                       \\ \hline
\multirow{6}{*}{OWASP}       & Consent-related issues                                                                                                     & All assets                                                                                                                                                                                                                                       \\ \cline{2-3} 
                             & Inability of user to access and modify data                                                                                & \begin{tabular}[c]{@{}l@{}}Map data, V2X information, Device information, User information,\\ Special categories of personal data, User preferences, Purchase information\end{tabular}                                                           \\ \cline{2-3} 
                             & Insufficient data breach response                                                                                          & \begin{tabular}[c]{@{}l@{}}Sensors data, Key and certificates, Map data, V2X information,\\ Device information, User information, Special categories of personal data,\\ User preferences, Purchase information\end{tabular}                     \\ \cline{2-3} 
                             & Misleading content                                                                                                         & \begin{tabular}[c]{@{}l@{}}Map data, V2X information, Device information, User information,\\ Special categories of personal data, User preferences\end{tabular}                                                                                 \\ \cline{2-3} 
                             & Secondary use                                                                                                              & All assets                                                                                                                                                                                                                                       \\ \cline{2-3} 
                             & Sharing, transfer or processing through 3rd party                                                                          & \begin{tabular}[c]{@{}l@{}}Sensors data, Key and certificates, Map data, V2X information,\\ Device information, User information, Special categories of personal data,\\ Driver's behaviour, User preferences, Purchase information\end{tabular} \\ \hline
\end{tabular}
\end{table*}

\end{landscape}

\bibliographystyle{unsrtnat}
\bibliography{references}  






\end{document}